\documentclass[twocolumn]{jpsj2}
\def\H{{\cal H}}
\def\JW{J_{\rm W}}
\def\JS{J_{\rm S}}
\def\Jint{J_{\rm int}}
\def\Hst{H_{\rm st}}
\def\chst{\chi_{\rm st}}
\def\v#1{\mib{#1}}
\newcommand{\aver}[1]{\left\langle {#1} \right\rangle}
\def\runauthor{Kazuo \textsc{Hida}}

\title{%
Field Induced Multiple Reentrant Quantum Phase Transitions in Randomly Dimerized Antiferromagnetic $S=1/2$ Heisenberg Chains }

\author{%
Kazuo \textsc{Hida}\thanks{E-mail: hida@phy.saitama-u.ac.jp}
}

\inst{%
Department~of~Physics,
 Faculty of Science,\\ Saitama University, Saitama, Saitama 338-8570
}

\recdate{\today}

\abst{%
The multiple reentrant quantum phase transitions in the $S=1/2$ antiferromagnetic Heisenberg chains  with random bond alternation in the magnetic field are investigated by the density matrix renormalization group method combined with the interchain mean field approximation. It is assumed that the odd-th bond is antiferromagnetic with strength $J$ and even-th bond can take the values ${\JS}$ and ${\JW}$ $ ({\JS} > J > {\JW} > 0)$ randomly with probability $p$ and $1-p$, respectively.  The pure version  ($p=0$ and $p=1$) of this model has a spin  gap but exhibits a field induced antiferromagnetism in the presence of interchain coupling if Zeeman energy due to the magnetic field exceeds the spin gap. For  $0 < p < 1$, the antiferromagnetism is induced by randomness at small field region where the ground state is disordered due to the spin gap in the pure case.  At the same time, this model exhibits  randomness induced plateaus at several values of magnetization. The antiferromagnetism is destroyed on the plateaus. As a consequence, we find a series of  reentrant quantum phase transitions between the transverse antiferromagnetic phases and disordered plateau phases with the increase of the magnetic field for moderate strength of interchain coupling. Above the main plateaus, the magnetization curve consists of a series of small plateaus and the jumps between them, It is also found that the antiferromagnetism is induced by infinitesimal interchain coupling at the jumps between the small plateaus. We conclude that this antiferromagnetism is supported by the mixing of low lying excited states by the staggered interchain mean field even though the spin correlation function is short ranged in the ground state of each chain.}

\kword{%
random quantum spin chain, DMRG, disorder induced order, field induced order, randomness induced plateau, reentrant phase transition
}
\begin{document}

\maketitle

\section{Introduction}
In the recent studies of one-dimensional quantum spin systems, the exotic quantum phases induced by the strong magnetic field  have been attracting broad interest. Among them, the field induced transverse antiferomagnetism  have been widely investigated in many experiemntal and theoretical studies.\cite{nikuni,honda,matsu,tanaka} If the magnetic field larger than the spin gap is applied to the spin gapped system, the single chain ground state becomes the Tomonaga-Luttinger liquid and the transverse antiferromagnetic order develops as soon as the weak interchain coupling is swtiched on. 

The disorder is another origin of order in the spin-gapped low dimensional quantum magnets\cite{manakanew,uchi,uchiprog,yasu1,kh1,vilar}.  In the presence of disorder, the spins in the nonmagnetic ground state revive and induce  so-called 'disorder induced order' even in the absence of the magnetic field. 

On the other hand, the possibility of disorder induced magnetization plateau is also predicted in a certain class of one-dimensional random quantum magnets.\cite{kh1,kh2,cabra,mike,noh} This corresponds to the spin gap state induced by disorder and magnetic field. 

Therefore, the effect of disorder on the quantum magnets in the magnetic field is twofold. Namely, it enhances the magnetic order by reviving the spins, while it suppresses the magnetic order by forming plateaus. In the present work, we investigate the competition between these two controversial aspects of randomness in the quasi-one dimensional quantum spin systems and the resulting multiple reentrant phase transitions between the transverse antiferromagnetic phases and disordered plateau phases. The similar problem has been discussed in the diluted dimer network system by Mikeska and coworkers\cite{mike} and in the  coupled random dimer network by Nohadani and coworkers\cite{noh}. 

This paper is organized as follows. In the next section, the model Hamiltonian is presented. The single chain magnetization curve is calculated in section 3. Within the interchain mean field approximation, we predict the multiple reentrant behavior with the increase of the magnetic field in section 4. The calculation of the spin-spin correlation function is presented in \S 5. Even in the non-plateau state, where the infinitesimal interchain coupling induces the transverse ordering, the correlation function of a single chain turned out to be short ranged. Based on these observations, the mechanism of the antiferromagnetism away from the plateau region is explained in \S 5.  The final section is devoted to summary and discussion.

\section{Model Hamiltonian}

As a candidate model in which the reentrant antiferromagnetism is expected, we investigate the quasi-one-dimensional random dimerized $S=1/2$ Heisenberg chain whose Hamiltonian is given by,
\begin{eqnarray}
\label{ham1}
\H &=& \sum_{j}\left\{\sum_{i=1}^{N/2}J  \v{S}_{2i-1,j} \v{S}_{2i,j} + \sum_{i=1}^{N/2}J_{ij}  \v{S}_{2i,j} \v{S}_{2i+1,j}\right\}\nonumber\\
&+&\sum_{i=1}^{N}\sum_{<j,j'>}J_{\rm int}  \v{S}_{i,j} \v{S}_{i,j'},
\end{eqnarray}
where  $J_i=\JS $ with probability $p$ and  $J_i=\JW $  with probability $1-p$. The interchain exchange coupling is denoted by $\Jint$. The spin operator $\v{S}_{i,j}$ denotes the spin on the $i$-th site of the $j$-th chain. The summation $\sum_{<j,j'>}$ is taken over all nearest neighbour pairs of chains. In the present work, we assume $\JS > J > \JW >0$. Similar model with ferromagnetic $\JW$ has been discussed\cite{kh1} related with the experimental materials.\cite{manakanew}

\begin{figure}
\centerline{\includegraphics[width=8cm]{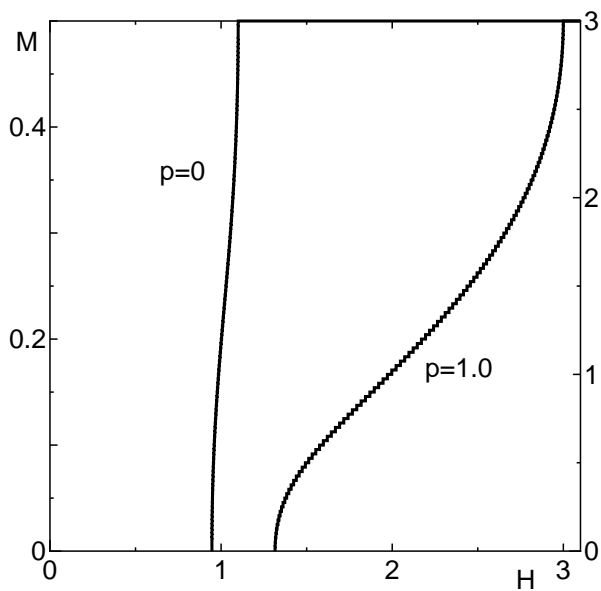}}
\caption{The magnetization curves of pure single chains with $p$=0 and 1.0. The chain length is $N=240$. }
\label{puremag}
\end{figure}

\begin{figure}
\centerline{\includegraphics[width=8cm]{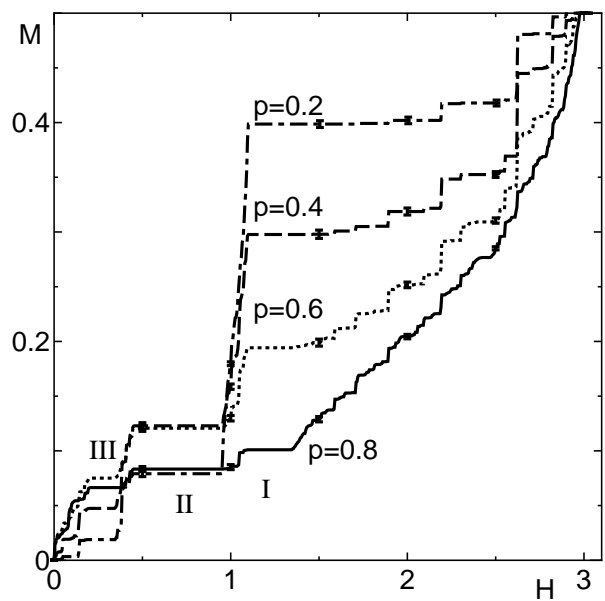}}
\caption{The magnetization curve of single chains for $p$=0.2, 0.4, 0.6 and 0.8. The chain length is $N=120$. The magnetization is measured for the middle 60 sites to reduce the boundary effect. The average is taken over 64 samples.  The error bars are shown only for selected points because otherwise the symbols are extremely dense.}
\label{magall}
\end{figure}

\section{Single Chain Magnetization curve}

The ground state magnetization curves of the single chain with $\Jint=0$ in (\ref{ham1}) is calculated using the DMRG method. The magnetization per site $M$ is defined by
\begin{equation}
M\equiv\frac{1}{N}\sum_{i=1}^N<S^z_i> 
\end{equation}
where the summation is taken over all spins in a single chain and the chain index $j$ is suppressed. The regular models with $p=0$ or $p=1$ has magnetization plateaus at $M=0$  which corresponds to the spin gap and at the saturation magnetization $M=M_{\rm s}\equiv 1/2$. However, it has no plateau with intermediate values of magnetization as shown in Fig. \ref{puremag}.

On the other hand, the magnetization curves for $p \ne 0, 1$ consist of a sequence of plateaus. Between them, the magnetization increases almost continuously. The typical example is shown in Fig. \ref{magall} for $\JS=2, \JW=0.1$ and $J=1$ for various values of $p$.

The main features of the magnetization curves can be understood using the cluster picture similar to that described in ref. \cite{kh1}.  With the increase of magnetic field, follow three large plateaus which are numbered I, II and III in Fig. \ref{magall}. Let us consider a cluster consisting of $q$ successive $\JS$-bonds and $q-1$ $J$-bonds in between. This is called  '$q$-cluster' as in ref. \cite{kh1}. The $2q$ spins in a cluster form a tightly bound  singlet cluster. The two spins connected to both ends of this cluster by $J$-bonds are almost free but weakly coupled  mediated by the quantum fluctuation within the strongly coupled cluster. Other spins form singlet dimers on the $J$-bonds if $\JW << J$.

On the plateau I with magnetization $M=(1-p)M_{\rm s}$, the spins which do not belong to the $q$-clusters are all polarized. On the plateau II, the end spins separated by 1-clusters with a single $\JS$-bond remain unpolarized. Similarly, on the plateau III, the end spins separated by  2-clusters also remain unpolarized and so on. These interpretation are confirmed by comparing with the magnetization process of a cluster consisting of a $q$-cluster and two additional end spins connected by $J$-bonds on both ends of the $q$-cluster. Lower plateaus due to the spins separated by longer $q$-clusters are not clearly identified within the present scale. The low field part of the magnetization curve reflects the singularity of the low energy excitation spectrum as described in ref. \citen{kh1}.

 Above the plateau I, the magnetization increases with series of plateaus and narrow continuous parts up to the saturation field. As $p$ increases, the width of the plateaus decrease and magnetization curve becomes almost continuous. 

\section{Effect of Interchain Exchange Interaction}

We treat the interchain coupling by the mean field approximation\cite{sca} assuming the transverse antiferromagnetic order as,
\begin{equation}
\aver{S^x_{i,j}} =\left\{\begin{array}{ll}
(-1)^{i}m & \Jint <0 \\
(-1)^{i}P_j m & \Jint >0 
\end{array}\right.
\end{equation}
For $\Jint > 0$, we assume that the two dimensional lattice of the chains is bipartite. The quantity $P_j$ is +1 if the site $j$ belongs to one of the sublattice and $-1$ if it belongs to the other. We thus have the interchain mean field Hamiltonian $H^{\rm IMF}$ for each chain as
\begin{eqnarray}
\label{ham2}
\H^{\rm IMF} &=& \sum_{i=1}^{N}J  \v{S}_{2i-1} \v{S}_{2i} + \sum_{i=1}^{N}J_i  \v{S}_{2i} \v{S}_{2i+1}\nonumber\\
&-&\Hst\sum_{i=1}^{N} (-1)^i{S}^x_{i},
\end{eqnarray}
with $\Hst=-z|J_{\rm int}|m$. 

In order to investigate the stability against transverse antiferromagnetic ordering, we fix the interchain mean field $\Hst$ and calculate the finite field staggered susceptibility $\chi (\Hst)=m(\Hst)/\Hst$ which tends to the staggered susceptibility $\chst$ as $\Hst$ tends to 0.  In general $\chi^{-1}_{\rm st}$ gives the minimum interchain coupling $\lambda=z|J_{\rm int}|$ which stabilizes the transverse ordering. If $\chst$ diverges, the transverse ordering takes place for the infinitesimal interchain coupling within the interchain mean field approximation.
\begin{figure}
\centerline{\includegraphics[width=8cm]{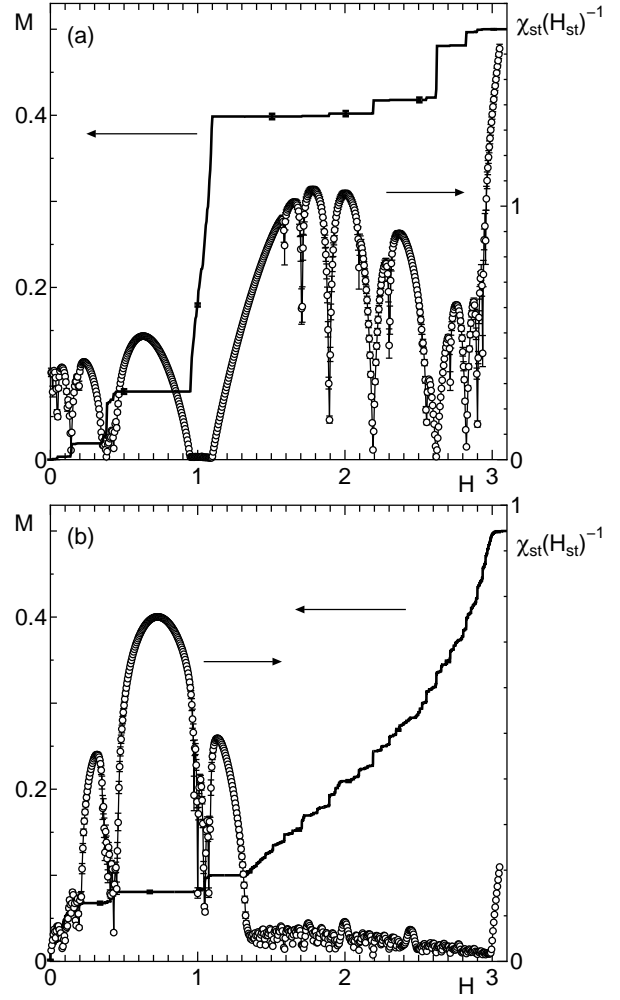}}
\caption{The $H$-dependence of $1/\chst(\Hst)$ with $\Hst=0.0005$ for (a) $p=0.2$ and (b) $p=0.8$. The magnetization curves are also shown for reference. The staggered magnetization $m$ is measured for the middle 60 sites to reduce the boundary effect. The chain length is $N=120$ and average is taken over 512 samples. The error bars are shown only for selected points.}
\label{a2a01chi}
\end{figure}
\begin{figure}
\centerline{\includegraphics[width=8cm]{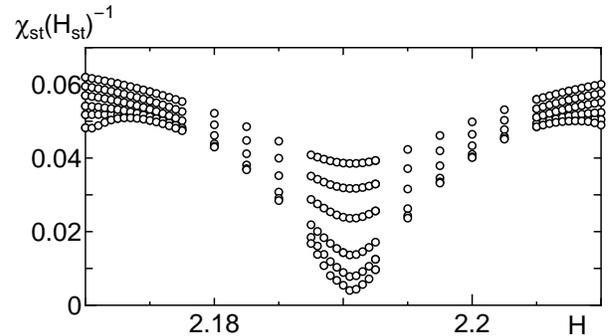}}
\caption{The $H$-dependence of $1/\chst(\Hst)$ for $2.17 \leq H \leq 2.21$. The values of the staggered field are $\Hst=0.002, 0.0015, 0.001, 0.0005, 0.00025$ and 0.0001 from top to bottom. The chain length is $N=480$ and measurement is done for the middle 240 sites.  The average is taken over 256 samples.The error bars are within the size of the symbols.}
\label{a2a01p08_ub3n48}
\end{figure}
\begin{figure}
\centerline{\includegraphics[width=8cm]{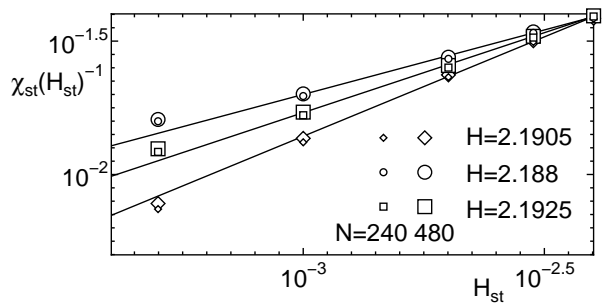}}
\caption{The $\Hst$-dependence of $\chst(\Hst)^{-1}$ around $H=2.1905$. The big symbols are for $N=480$ and small ones are for $N=240$. The solid lines are power law extrapolation from $\Hst=0.0005, 0.001, 0.0015$ and 0.002 for $N=480$. The average is taken over 256 samples.}
\label{a2a01p083phchi}
\end{figure}
\begin{figure}
\centerline{\includegraphics[width=8cm]{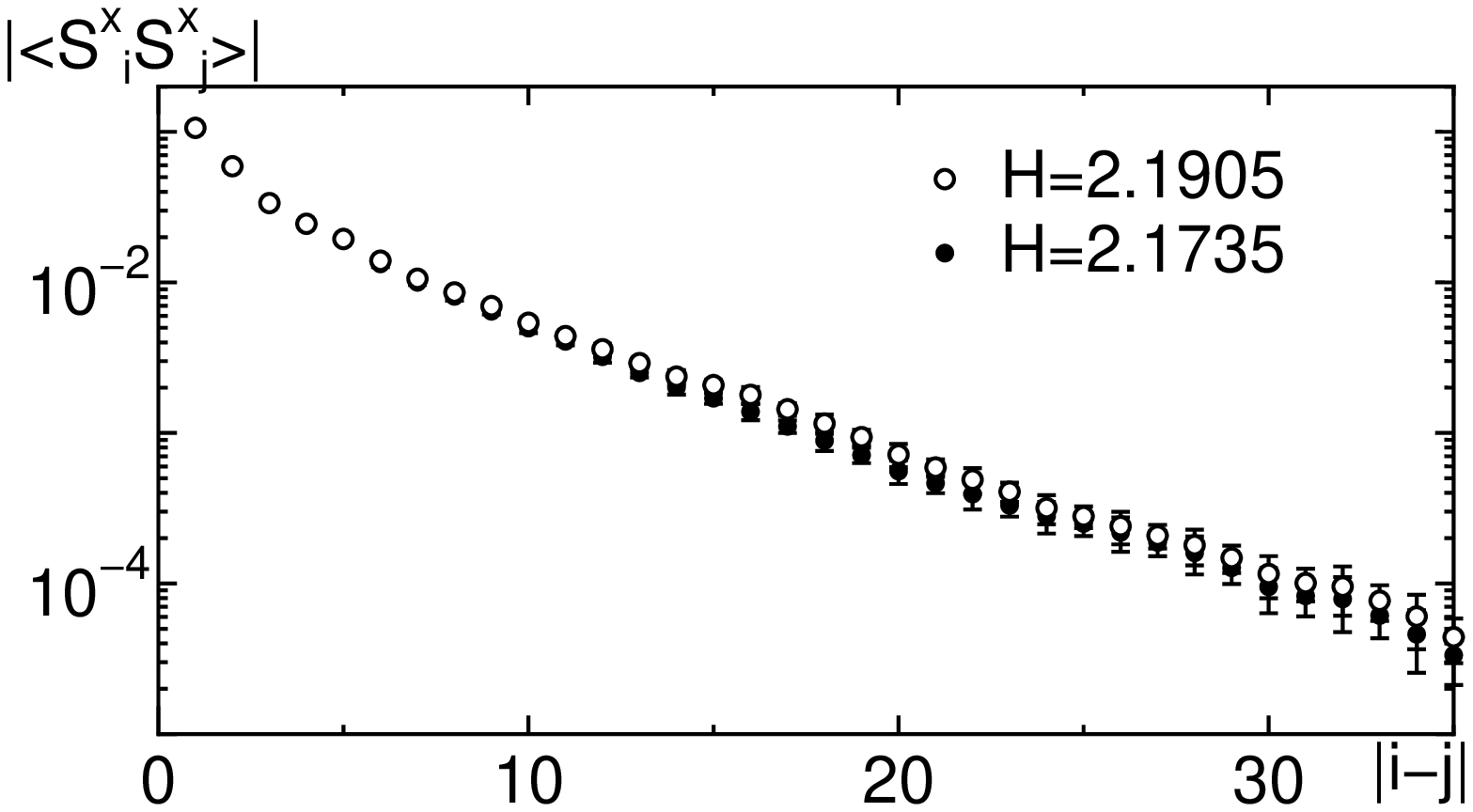}}
\caption{The transverse correlation function $|\aver{S^x_iS^x_j}|$ at $H=2.1905$($\circ$) where $\chst(\Hst\rightarrow 0)=0$ and at $H=2.1735$($\bullet$) where $\chst(\Hst\rightarrow 0)=finite$.  The average is taken over 512 samples.}
\label{a2a01p08b3merc}
\end{figure}
\begin{figure}
\centerline{\includegraphics[width=8cm]{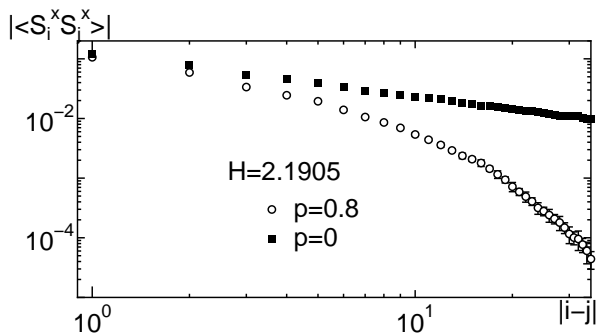}}
\caption{The {\rm log}-{\rm log} plot of the correlation function with $p=0.8$ (open circles) and  $p=1$(filled squares) at $H=2.1905$.}
\label{a2a01p08b3merc0}
\end{figure}
\begin{figure}
\centerline{\includegraphics[width=70mm]{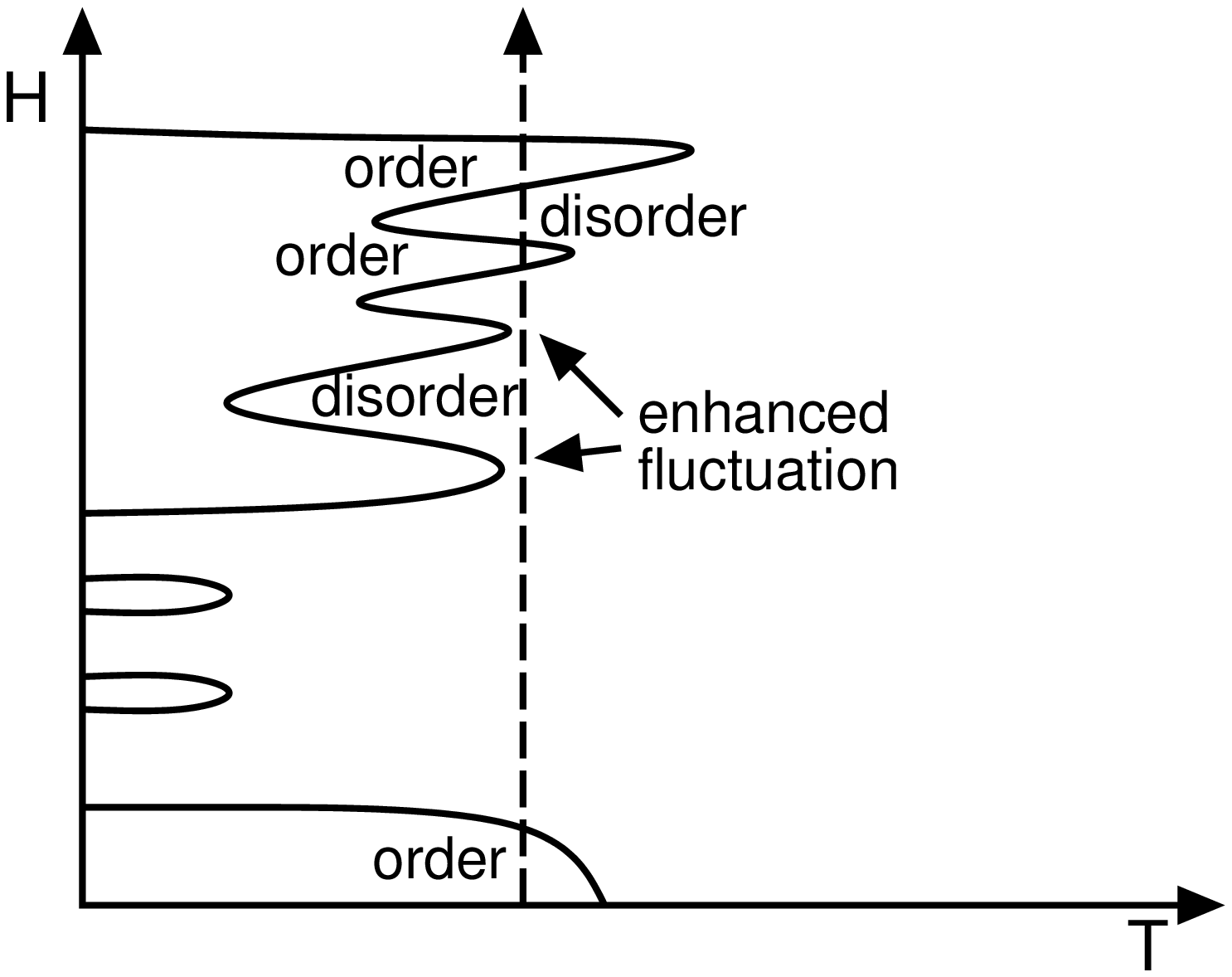}}
\caption{Schematic finite temperature phase diagram on the $H-T$-plane.}
\label{schema}
\end{figure}

Fig. \ref{a2a01chi} shows the magnetic field dependence of $\chst^{-1}(\Hst)$  with $\Hst=0.0005$ for $p=0.2$ and $p=0.8$ as representatives of the small $p$ and large $p$ cases. The magnetization curves in the absence of staggered field is also presented. The $H$-$\chi^{-1}(\Hst)$ curve has multiple maxima, which clearly shows that multiple reentrant behavior takes place for finite interchain coupling in the ground state.  It should be noted that the $H$-$\chi^{-1}(\Hst)$ curves are insensitive to the value of $\Hst$ around these maxima. However, around the dips and minima, the values of $\chi(\Hst)$ have significant $\Hst$-dependence. 

For $p=0.8$,  $\chst^{-1}(\Hst)$ remains significantly small above the main plateaus compared to the peak values on the main plateaus.  Therefore we expect no disordered phase for moderate values of interchain coupling in this region  where the magnetization curve appears almost continuous. Even away from the main plateaus, however, the magnetization curve shows a series of  small plateaus and narrow continual parts between them. Correspondingly, $\chst^{-1}(\Hst)$ tends to a finite value as $\Hst \rightarrow 0$ on these plateaus. The detailed features of such behavior is shown in Fig.\ref{a2a01p08_ub3n48} for $2.17 \leq H \leq 2.21$ as a representative. On these small plateaus,  $\chst^{-1}(\Hst)$ clearly tends to  small finite values as shown in Fig. \ref{a2a01p08_ub3n48} around $H=2.173$ and $H=2.208$. On the other hand, in the true off-plateau state, $\chst$ tends to zero suggesting the divergence of $\chst$. In this case, the transverse antiferromagnetic order is stabilized in the presence of infinitesimal interchain interaction. In Fig. \ref{a2a01p08_ub3n48}, such behavior is observed at $H=2.1905$. To investigate this behavior in more detail, we present the $\Hst$-dependence of $\chst$ in Fig. \ref{a2a01p083phchi} at $H=2.1925, 2.1905$ and 2.188 using the data for $N=$480. Only 0.2\% deviation from $H=2.1905$ causes a clear upturn of $\chst^{-1}(\Hst)$. Although the small size dependence is present, the data for $N=240$ also shows the similar behavior as plotted in smaller symbols. Therefore we expect this is not the finite size effect but is an essential feature of the present system in the thermodynamic limit. Similar behavior is observed for other values of $H$ above the main plateaus. Therefore we conclude that the antiferromagnetic order is stabilized by infinitesimal interchain coupling only within a narrow region where magnetization increases continously between the successive small pleteaus. 

\section{Correlation Functions}
In order to get more insight into the nature of each state based on the properties of single chains, we have also investigated the spin-spin correlation function $\aver{S^x_iS^x_j}$ as a function of $|i-j|$. Figure \ref{a2a01p08b3merc} shows $|\aver{S^x_iS^x_j}|$ for $H=2.1905$($\circ$) where $\chst(\Hst)$ tend to zero. Even in this case, the spin-spin correlation function is short ranged. Actually, the behavior of the corelation function is almost the same as that for $H=2.1735$($\bullet$) where $\chst(\Hst)$ tends to a small but finite value. In Fig.\ref{a2a01p08b3merc0}, the {\rm log-log} plot of the same correlation function is compared with that for the regular chain with $p=1$ at $H=2.1905$. It is clear that the rapid decrease of the correlation is distinct from the power decay for the regular chain.

This can be understood in the following way. In the off-plateau region, the continuum of the low energy excited states  pile up on the ground state. In many of these excited states, the spins which are not correlated in the ground state are correlated. The staggered transverse magnetic field mixes up these excited states and leads to the divergent staggered susceptibility. In this case, the long range transverse order can be stablized with infinitesimal interchain coupling even though the spin correlation is short ranged in the ground state. On the other hand, in the plateau state there exists no low energy excited states which supports the long range order with small interchain coupling.

\section{Summary and Discussion}

The transverse magnetic ordering in the ground state of the random quantum Heisenberg chain is investigated using the density matrix renormalization group and the interchain mean field approximation. It is predicted that the multiple reentrant behavior takes place between the disordered plateau phases and transverse antiferromagnetic ordered phases. This is in contrast to the case of random dimer networks discussed by Mikeska and coworkers\cite{mike} and  Nohadani and coworkers\cite{noh} for which the reentrant transisition takes place only once.

It is also pointed out that even in the non-plateau regime the spin-spin correlation of the single chain is short ranged. Nevertheless, the long range order is established with infinitesimal interchain interaction with the help of the excited states which pile up near the ground state and are mixed up by the interchain staggered mean field.

In this work we concentrated on the ground state phase transition. Nevertheless, the reentrant behavior should survive even at finite temperatures. The transition temperature should be high between the plateau region and low or zero on the plateau region as depicted in Fig. \ref{schema} schematically. This behavior manifests itself as anomalous behavior even above the transition temperature. Therefore, if we increase the magnetic field with fixed temperature and passes near the ordered phase, the transverse spin fluctuation would be strongly enhanced. 

We expect the present type of reentrant behavior is universal in the random quantum spin systems in which the singlet dimer formation is randomly perturbed to produce the local almost free spins. In contrast to the random dimer network systems in which the dilution produces the isolated spins, the free spins in the present system are produced by the random competition of two different dimer interactions $\JS$ and $J$ each of which prefers different dimer configuration. This is the origin of more complicated structure of the phase diagram. Thus we expect the reentrant behavior of due to similar mechanism in a variety of systems. 

 The computation in this work has been done using the facilities of the Supercomputer Center, Institute for Solid State Physics, University of Tokyo and the Information Processing Center, Saitama University.  This work is supported by a Grant-in-Aid for Scientific Research from the Ministry of Education, Culture, Sports, Science and Technology, Japan.

\end{document}